\documentclass[11pt, notitlepage]{article}
\pdfoutput=1

\usepackage{float}
\usepackage{subfig}
\usepackage{graphicx}
\usepackage[T1]{fontenc}
\usepackage[utf8]{inputenc}
\usepackage{authblk}
\usepackage{amsmath,bm}
\usepackage{amssymb}
\usepackage{amsmath}
\usepackage{enumitem}
\usepackage{mathtools}
\usepackage[left=2cm,right=2cm,top=2cm,bottom=2cm]{geometry}

\renewcommand\footnotemark{}
\begin{document}

\title{\bf{Asymptotically Weyl-Invariant Gravity}}
\author{Daniel Coumbe}
\thanks{E-mail: daniel.coumbe@nbi.ku.dk}
\affil{\small{\emph{The Niels Bohr Institute, Copenhagen University}\\ \emph{Blegdamsvej 17, DK-2100 Copenhagen Ø, Denmark}}}
\date{}
\maketitle


\begin{abstract}

  We propose a novel theory of gravity that by construction is renormalizable, evades Ostragadsky's no-go theorem, is locally scale-invariant in the high-energy limit, and equivalent to general relativity in the low-energy limit. The theory is defined by a pure $f(R)=R^{n}$ action in the Palatini formalism, where the dimensionless exponent $n$ runs from a value of two in the high-energy limit to one in the low-energy limit. We show that the proposed model contains no obvious cosmological curvature singularities. The viability of the proposed model is qualitatively assessed using several key criteria. 

\vspace{0.25cm}

\end{abstract}



\begin{section}{Introduction}

General relativity is theoretically beautiful, experimentally successful, and defines our best current description of gravity. Its key equations follow almost inevitably from a single symmetry principle and yield predictions that agree with experiment over a vast range of distance scales~\cite{Will:2014kxa}. 

Yet, general relativity is almost certainly incomplete. One major problem is that it appears to be incompatible with quantum field theory at high energies. Although general relativity has been successfully formulated as an effective quantum field theory at low energies, new divergences appear at each order in the perturbative expansion, leading to a complete loss of predictivity at high energies. Gravity is said to be perturbatively non-renormalizable, a fact demonstrated by explicit calculation at the one-loop level including matter content~\cite{'tHooft:1974bx}, and at the two-loop level without matter~\cite{Goroff:1985th}. This behaviour stems from the fact that the gravitational coupling introduces a length scale into the theory so that higher-order corrections in a perturbative expansion come with ever-increasing powers of the cut-off scale. 

Higher-order actions have been explored as a possible solution to this problem~\cite{Belenchia:2016bvb}. Explicit calculations show that Lagrangians quadratic in the curvature tensor are renormalizable~\cite{Stelle:1977nt}. However, quadratic theories of gravity are not hailed as \emph{the} theory of quantum gravity because they are not typically unitary, often containing unphysical ghost modes~\cite{Stelle:1977nt}. Higher-order theories are also generally unstable, violating a powerful no-go theorem first proposed by Ostragadsky~\cite{Ostrogradsky}. Ostrogradsky's theorem proves that there is a fatal linear instability in any Hamiltonian associated with Lagrangians which depend on two or more time derivatives. This result helps explain the otherwise mysterious fact that the fundamental laws of physics seem to include at most two time derivatives~\cite{Woodard:2006nt}. The only higher-order theories that are both stable and unitary are $f(R)$ theories, in which the Lagrangian is a general function $f$ of the Ricci scalar $R$~\cite{Sotiriou:2008rp}. 

Since $f(R)$ gravity is not a single theory but a potentially infinite set of theories, one for each particular function $f(R)$, we must identify a principle capable of selecting a specific function $f(R)$. We contend that this principle should be local scale invariance. One reason for this is that scale-invariant theories of gravity are gauge theories~\cite{Wesson,Lasenby:2015dba}. Since the electromagnetic, strong and weak interactions are all local gauge theories it is natural to posit that gravity is too. Such a gauge theory of gravity would open the possibility of unification with the other three interactions. Modern gauge theory actually originates from the attempt to make gravity locally scale-invariant, as first proposed by Hermann Weyl~\cite{Weyl:1918ib}. Weyl reasoned that all length measurements are local comparisons. For example, measuring the length of a rod requires a local comparison with some standard unit of length, say a meter stick, with the result being the dimensionless ratio of their lengths. Empirically, this is what it means to say something has a length of $x$ units. Repeating the measurement at a different spacetime point, where the metric is rescaled by some factor, must yield the same dimensionless ratio since the rod and standard length are both equally rescaled. Weyl reasoned that local changes of scale should therefore be physically unobservable, and that the laws of physics must reflect this by being invariant under local rescalings of the metric $g_{\mu\nu} \to g_{\mu\nu}\Omega^{2}(x)$, where $\Omega(x)$ is the factor by which $g_{\mu\nu}$ is rescaled at the point $x$. Mathematically, Weyl pointed out that Riemannian geometry assumes the magnitude of two vectors can be compared at arbitrarily separated points, contrary to a purely infinitesimal geometry in which only local measurements have any meaning~\cite{Weyl:1918ib}. \emph{A priori} there is no fundamental reason for this assumption~\cite{Weyl:1918ib,Straumann:1996ji}. 

Although an initial admirer of Weyl's proposal, Einstein later pointed out a fatal flaw in his idea~\cite{Weyl:1918ib}. The spacetime interval in this scenario is in general path-dependent. Einstein argued that this implies atomic spectral lines can depend on the particular path an element takes through spacetime~\cite{Weyl:1918ib}. Since this prediction has not been observed within current experimental limits it seems to be empirically ruled out. More obviously, our universe cannot possibly be scale-invariant at low energies due to the existence of particles of non-zero rest mass that introduce an associated length scale, hence breaking exact scale invariance. For example, the local scale invariance of the standard model of particle physics is broken by the Higgs mass term~\cite{Bars:2013yba}.  

Yet, an exciting possibility remains. It is conceivable that our universe asymptotically approaches exact local scale invariance in the high-energy limit. This has the practical advantage that the spacetime interval will remain effectively path-independent at all but the most extreme energy scales, safely pushing Einstein's criticism of Weyl's proposal beyond current experiment limits. There are also numerous physical reasons to expect this behaviour. Firstly, at very high energies particle masses become negligable since in the ultra-relativistic limit of special relativity the massless dispersion relation becomes increasingly accurate. It therefore seems natural to expect a high-energy theory to lack any explicit mass scales~\cite{Aalbers:2013jjv}. Secondly, any renormalizable quantum field theory must scale in the same way as a conformal field theory at high energies~\cite{Banks:2010tj,Shomer:2007vq}. Thus, renormalization only requires exact local scale invariance in the high energy limit~\cite{Shomer:2007vq,Smolin:1979uz}. Thirdly, measurements by WMAP and the Planck satellite shows that the spectrum of primordial fluctuations is near scale-invariant~\cite{Akrami:2018odb}. This simplicity may be a hint of an underlying symmetry of nature, rather than a very special set of initial conditions in the early universe. Finally, conformal symmetry seems deeply connected with gravity as exemplified by the AdS/CFT correspondence~\cite{Maldacena:1997re}. 

We may also indulge in a more aesthetic argument. Symmetry principles have proven remarkably successful in revealing the fundamental laws of nature. In particular, the progression from Newtonian mechanics to special and general relativity can be viewed as a successive increase in the symmetry of spacetime. Can this trend be continued? Under a plausible set of assumptions, the Coleman-Mandula theorem states that the maximal symmetry spacetime can have is conformal symmetry~\cite{Coleman:1967ad,FioresiBook}. Conformal transformations are nothing but a special type of Weyl transformation, in which the rescaled metric must remain diffeomorphic to the original~\cite{Karananas:2015ioa,Farnsworth:2017tbz}. Maximising the symmetry of spacetime should therefore severely constrain, and possible even uniquely select, the laws of nature.


The only $f(R)$ theory that is globally scale-invariant is pure $R^{2}$ gravity, as well as being the only unitary theory that is purely quadratic in the curvature tensor~\cite{Kehagias:2015ata,Kounnas:2014gda}. Pure $R^{2}$ gravity is invariant under the global scale transformation $g_{\mu\nu} \rightarrow g_{\mu\nu}\Omega^{2}$, where $\Omega$ is a constant~\cite{Kehagias:2015ata}. However, it is not invariant under a local scale (Weyl) transformation 

\begin{equation}\label{weyl1}
g_{\mu\nu} \rightarrow g_{\mu\nu}\Omega^{2}(x), \qquad g^{\mu\nu} \rightarrow g^{\mu\nu}\Omega^{-2}(x),
\end{equation}

\noindent where $\Omega(x)$ is a point-dependent function. This is because in four dimensions $R^{2}$ transforms under~(\ref{weyl1}) as~\cite{Dabrowski:2008kx}

\begin{equation}
R^{2} \rightarrow \frac{1}{\Omega^{4}(x)}\left[R^{2}+36\frac{\left(\square \Omega(x)\right)^{2}}{\Omega^{2}(x)} - 12R\frac{\square \Omega(x)}{\Omega(x)}\right],
\end{equation}

\noindent whereas the determinant of the metric transforms under~(\ref{weyl1}) according to $\sqrt{-g} \rightarrow \Omega^{4}(x)\sqrt{-g}$~\cite{Dabrowski:2008kx}. Thus, an action with $f(R)=R^{2}$ is not locally scale-invariant because its Lagrange density $R^{2}\sqrt{-g}$ is not, and therefore it cannot fulfill the symmetry we seek at high energies. In fact, there is no metric $f(R)$ theory that is Weyl invariant.\interfootnotelinepenalty=10000 \footnote{\scriptsize For non-$f(R)$ conformally-invariant theories of gravity see~\cite{Mannheim:2011ds,tHooft:2016uxd}.} Weyl symmetry is already proving to be a powerful constraint. 

Fortunately, there is more than one type of $f(R)$ theory; actually there are three~\cite{Sotiriou:2008rp}. The first is referred to as the metric formalism, in which it is assumed that the affine connection depends on the metric in a unique way via the Levi-Civita connection, as in standard general relativity. This is the formulation already discussed. The second is referred to as the Palatini formalism, which generalises the metric formalism by relaxing the assumption that the connection must depend on the metric. The third is known as the metric-affine formalism, which generalises the Palatini formalism by dropping the implicit assumption that the matter action is independent of the connection. As shown above, metric $f(R)$ gravity does not permit a Weyl-invariant action. The metric-affine formalism is not a metric theory, meaning the energy tensor is not divergence-free and diffeomorphism invariance is almost certainly broken~\cite{Sotiriou:2008rp}.\interfootnotelinepenalty=10000 \footnote{\scriptsize There is still some debate on the exact interpretation of this result~\cite{Sotiriou:2008rp}.} This leaves us to consider the Palatini formulation of $f(R)$ gravity.  

General relativity assumes that the affine connection uniquely depends on the metric via the Levi-Civita connection. However, in differential geometry the metric and affine connection are generally independent objects. The metric defines distances and angles on the manifold, while the affine connection defines the intrinsic curvature of the manifold via parallel transport. \emph{A priori} there is no reason for assuming the affine connection must depend on the metric. The Palatini formalism, which was first proposed by Einstein~\cite{Ferraris1982}, is a generalisation of general relativity in which this assumption is not made. Namely, in the Palatini formalism the connection $\Gamma^{\lambda}_{\mu\nu}$ is not assumed to depend on the metric $g_{\mu\nu}$. In addition to being theoretically simpler, in that it makes fewer assumptions, the Palatini formalism has spawned several important advances including the ADM formulation~\cite{Arnowitt:1962hi}, supergravity~\cite{Ortin:2015hya}, quantization via Ashtekar variables~\cite{Ashtekar:2004eh}, and others~\cite{Deser:1969wk}. Moreover, the Palatini formulation is closer in spirit to the ideas underlying general relativity. Since the metric formalism assumes the connection depends on the metric it also assumes the action contains second-order derivatives of the metric. This fact necessitates adding a surface term related to the extrinsic curvature, which requires a background metric in which to embed the spacetime. However, in Palatini gravity the action contains no metric derivatives and only first order derivatives of the connection~\cite{Olmo:2011uz}. Although it still requires surface terms they do not refer to any background spacetime, thus being closer to the original ethos of background independence. Unlike the metric-affine formalism, the Palatini formalism is a metric theory, in the strict sense that it satisfies the metric postulates, and is explicitly diffeomorphism invariant~\cite{Sotiriou:2008rp}.

Palatini gravity has another principle advantage, one that is particularly relevant to this work; it enables a simple Weyl-invariant action. Since the connection is independent of the metric in the Palatini formalism, the Weyl transformation~(\ref{weyl1}) leaves the connection invariant. The Ricci scalar $\mathcal{R}$ in the Palatini formalism is defined by

\begin{equation}
\mathcal{R}\vcentcolon = g^{\mu\nu}\mathcal{R}_{\mu\nu}=g^{\mu\nu}\left(\partial_{\rho}\Gamma^{\rho}_{\nu\mu} - \partial_{\nu}\Gamma^{\rho}_{\rho\mu} + \Gamma^{\rho}_{\rho\lambda}\Gamma^{\lambda}_{\nu\mu} - \Gamma^{\rho}_{\nu\lambda}\Gamma^{\lambda}_{\rho\mu}\right),
\end{equation}

\noindent and so under a Weyl transformation $\mathcal{R}^{2}$ transforms simply via $\mathcal{R}^{2} \rightarrow \Omega^{-4}(x)\mathcal{R}^{2}$. Therefore, pure $\mathcal{R}^{2}$ gravity in the Palatini formalism is Weyl invariant, because its Lagrange density $\mathcal{R}^{2}\sqrt{-g}$ is unchanged by a Weyl transformation~\cite{Borowiec:1996kg,Edery:2019txq}. Hence, $f(\mathcal{R})=\mathcal{R}^{2}$ possesses the symmetry we desire in the high-energy limit. 

However, in order to be consistent with experimental tests of general relativity and reproduce our scale-dependent diffeomorphism invariant universe we must recover the standard Einstein-Hilbert action in the low-energy limit. Remarkably, $f(\mathcal{R})=\mathcal{R}$ gravity in the Palatini formalism is identical to standard general relativity, since in this specific case the Levi–Civita connection turns out to be a dynamical result of the equations of motion~\cite{Sotiriou:2008rp,WaldBook}.\interfootnotelinepenalty=10000 \footnote{\scriptsize This equivalence is specific to actions linear in the curvature tensor and does not hold for higher-order theories~\cite{BeltranJimenez:2017vop,Sotiriou:2008rp}.} Therefore, it is sufficient to require our action reduces to $f(\mathcal{R})=\mathcal{R}$ in the low energy limit.

\end{section}


\begin{section}{The model}\label{action}
\begin{subsection}{Action}

Based on this discussion we propose the following action for pure gravity

\begin{equation}\label{a1}
\mathcal{S}_{p}=\frac{1}{\kappa} \int \mathcal{R}^{h(k)}\sqrt{-g}d^{4}x,\qquad h(k)\to
\begin{cases}
      1, & \text{in the IR}, \\
      2, & \text{in the UV},
    \end{cases}
\end{equation}

\noindent where $\kappa$ is the gravitational coupling constant, $\mathcal{R}$ is the Ricci scalar in the Palatini formalism, $h(k)$ is a dimensionless function of the energy scale $k$ and $g$ is the determinant of the metric tensor. The scale $k$ is a classical coarse-graining scale, which can be thought of as the resolving power of a hypothetical microscope~\cite{Reuter:2005bb,Reuter:2006zq}.\interfootnotelinepenalty=10000 \footnote{\scriptsize One could also write an equivalent action which is purely a function of curvature, namely $f(\mathcal{R})=R^{h(\mathcal{R})}$, where $h(\mathcal{R})$ is dimensionless and $h(\mathcal{R})\to 1$ for small $R$ and $h(\mathcal{R})\to 2$ for large $R$. However, this somewhat complicates matters when considering partial derivatives of $f(\mathcal{R})$ with respect to $\mathcal{R}$ in later sections.} The abbreviations IR and UV refer to the infrared (low-energy limit) and ultraviolet (high-energy limit), respectively. The action of Eq.~(\ref{a1}) remains invariant under arbitrary differential coordinate transformations since we are only considering dimensionless powers of the Ricci scalar. 

Since in $4$-dimensional spacetime $\mathcal{R}^{h(k)}$ has a canonical mass dimension $[\mathcal{R}^{h(k)}]=2h(k)$ and $[\sqrt{-g}]=-4$, a dimensionless action requires $[\kappa]=4-2h(k)$. In the high-energy limit $h(k)\rightarrow 2$ and so $[\kappa]=0$; hence the gravitational coupling becomes dimensionless. The theory that this action defines is by construction renormalizable, unitary, stable, locally scale-invariant in the high-energy limit and equivalent to general relativity in the low-energy limit. We refer to this theory as asymptotically Weyl-invariant gravity (AWIG). 

The total gravitational action $\mathcal{S}$ can be written as the sum of the pure action $\mathcal{S}_{p}$ and the matter action $\mathcal{S}_{m}$, namely

\begin{equation}\label{a2}                                                                                                                                                                   
\mathcal{S}=\frac{1}{\kappa} \int \mathcal{R}^{h(k)}\sqrt{-g}d^{4}x + \mathcal{S}_{m}\left(g_{\mu\nu},\psi\right),                                                                           
\end{equation}                                                                                                                                                                               
\noindent where in the Palatini formalism $\mathcal{S}_{m}$ depends only on the metric $g_{\mu\nu}$ and the matter fields $\psi$, and not on the connection~\cite{Sotiriou:2008rp}.

Hereafter, we denote the function specific to our model by $F(\mathcal{R})\vcentcolon= \mathcal{R}^{h(k)}$ and denote an arbitrary function of the Ricci scalar in the Palatini formalism via the lower case notation $f(\mathcal{R})$. We denote the first derivative of $F(\mathcal{R})$ with respect to $\mathcal{R}$ by $F'(\mathcal{R})$, and the second derivative of $F(\mathcal{R})$ with respect to $\mathcal{R}$ by $F''(\mathcal{R})$. For future reference 

\begin{equation}\label{deriv0}
F(\mathcal{R})=\mathcal{R}^{h(k)},
\end{equation}

\begin{equation}\label{deriv1}
F'(\mathcal{R})=\mathcal{R}^{h(k)-1}\left(h(k)\right),
\end{equation}

\noindent and

\begin{equation}\label{deriv2}
F''(\mathcal{R})=\mathcal{R}^{h(k)-2}h(k)\left(h(k)-1\right).
\end{equation}

\end{subsection}


\begin{subsection}{Field equations}

In the metric formalism, the field equations can be obtained by varying the action with respect to the inverse metric tensor. However, in the Palatini formalism, the connection is independent of the metric and so one must vary the action separately with respect to the metric and the connection, yielding two equations. Varying with respect to the metric gives the field equations

\begin{equation}\label{Fieldeqns1}
F'(\mathcal{R})\mathcal{R}_{\mu\nu} -\frac{1}{2}F(\mathcal{R})g_{\mu\nu} =\kappa T_{\mu\nu},
\end{equation}

\noindent while varying with respect to the connection and performing some additional manipulations yields 

\begin{equation}\label{constraint1}
\nabla_{\lambda}\left(\sqrt{-g}F'(\mathcal{R})g^{\mu\nu}\right)=0,
\end{equation}

\noindent where we have used the relation $\delta \mathcal{R}_{\mu\nu}=\tilde{\nabla}_{\lambda}\delta \Gamma^{\lambda}_{\mu\nu}-\tilde{\nabla}_{\nu}\delta \Gamma^{\lambda}_{\mu\lambda}$, with the covariant derivative $\tilde{\nabla}_{\lambda}$ defined via a metric independent connection $\Gamma^{\lambda}_{\mu\nu}$~\cite{Sotiriou:2008rp}. The energy-momentum tensor $T_{\mu\nu}$ is defined in the usual way as
                                                                                                                                                 
\begin{equation}                                                                                                                                                                            
T_{\mu\nu}= \frac{-2}{\sqrt{-g}}\frac{\delta {S}_{M}}{\delta g^{\mu\nu}}.
\end{equation}

\noindent We may further simplify~(\ref{Fieldeqns1}) by taking its trace, giving

\begin{equation}\label{traceFE}
F'(\mathcal{R}) \mathcal{R}- 2F(\mathcal{R})=\kappa T,
\end{equation}

\noindent which provides an expression that will prove to be useful in later sections. The vacuum field equations are trivially obtained by setting $T=0$ in~(\ref{traceFE}), and using~(\ref{deriv1}) we have

\begin{equation}\label{VactraceFE}
F'(\mathcal{R}) \mathcal{R}- 2F(\mathcal{R})=\mathcal{R}^{h(k)}h(k)-2\mathcal{R}^{h(k)}=0.
\end{equation} 

The symmetry of local scale invariance is only exact in the high energy limit $h(k)\to 2$, for which $T=0$ identically. So in this case, gravity can only couple to conformally invariant matter, such as electromagnetic radiation. As the energy scale is lowered, gravity can couple to non-conformal matter, which breaks local scale invariance due to the introduction of a length scale associated with particles of non-zero rest mass.

\end{subsection}

\end{section}


\begin{section}{Application to curvature singularities}



General relativity is thought to predict its own breakdown due to the existence of curvature singularities. One must be careful that the singularity in question is not an artefact of the particular coordinate system used. For this reason, scalar measures of curvature are useful for identifying singularities, since they are coordinate invariant. 

The simplest curvature invariant is the Ricci scalar $R$. According to a well-known procedure~\cite{Magnano:1993bd,Higgs:1959jua,Barrow:1988xh} we may convert between the Jordan and Einstein frames via a conformal transformation of the metric

\begin{equation}\label{cs1}
g_{\mu\nu} \rightarrow \tilde{g}_{\mu\nu}=F'(\mathcal{R}) g_{\mu\nu}= \Omega_{k}^{2}(x) g_{\mu\nu}.
\end{equation}

\noindent Note that we have added the subscript $k$ to the conformal factor $\Omega_{k}^{2}(x)$ to signify its $k$ dependence. The curvature scalar $\mathcal{R}$ in the Palatini formalism transforms under~(\ref{cs1}) according to 

\begin{equation}\label{Rtrans}
\mathcal{R} \to \tilde{\mathcal{R}}=\frac{\mathcal{R}}{\Omega_{k}^{2}(x)},
\end{equation}

\noindent where $\tilde{\mathcal{R}}$ denotes the conformally transformed curvature scalar. Applying~(\ref{cs1}) and~(\ref{deriv1}) to~(\ref{Rtrans}), we have that

\begin{equation}
\mathcal{R}\rightarrow \tilde{\mathcal{R}} =\frac{\mathcal{R}}{F'(\mathcal{R})}=\frac{\mathcal{R}}{\mathcal{R}^{h(k)-1}h(k)}=
\begin{cases}
      \mathcal{R}, & \text{for}\ h(k) \rightarrow 1,\\
      \frac{1}{2}, & \text{for}\ h(k) \rightarrow 2.
\end{cases}
\end{equation}

\noindent The next simplest curvature invariant is $\mathcal{R}^{2}$, which transforms under~(\ref{cs1}) according to 

\begin{equation}
\mathcal{R}^{2}\rightarrow \tilde{\mathcal{R}}^{2}=\frac{\mathcal{R}^{2}}{\left(F'(\mathcal{R})\right)^{2}}=\frac{\mathcal{R}^{2}}{\mathcal{R}^{2h(k)-2}h^{2}(k)}=
\begin{cases}
      \mathcal{R}^{2}, & \text{for}\ h(k) \rightarrow 1,\\
      \frac{1}{4}, & \text{for}\ h(k) \rightarrow 2,
\end{cases}
\end{equation}

\noindent where we have used~(\ref{deriv2}). In fact, for any power $n$ of the Ricci scalar $\mathcal{R}$ the result will be a constant in the high-energy limit. Specifically, it is straightforward to show that $\mathcal{R}^{n}\to 1/2^{n}$ for $h(k) \rightarrow 2$. 

We now move on to the second-order curvature invariant involving the Ricci tensor, namely $\mathcal{R}_{\mu\nu}\mathcal{R}^{\mu\nu}$. Since in the Palatini formalism the connection $\Gamma^{\lambda}_{\mu\nu}$ is \emph{a priori} independent of the metric $g_{\mu\nu}$ the Ricci tensor

\begin{equation}
\mathcal{R}_{\mu\nu}=\partial_{\rho}\Gamma^{\rho}_{\nu\mu} - \partial_{\nu}\Gamma^{\rho}_{\rho\mu} + \Gamma^{\rho}_{\rho\lambda}\Gamma^{\lambda}_{\nu\mu} - \Gamma^{\rho}_{\nu\lambda}\Gamma^{\lambda}_{\rho\mu}
\end{equation}

\noindent will remain invariant under the metric transformation~(\ref{weyl1}). However, the Ricci tensor with upper indices is given by $\mathcal{R}^{\mu\nu}=g^{\mu \rho}g^{\nu \sigma}\mathcal{R}_{\rho \sigma}$, and so it will transform under~(\ref{weyl1}) as $\mathcal{R}^{\mu\nu} \to \mathcal{R}^{\mu\nu}\Omega_{k}^{-4}(x)$. Therefore, we have 

\begin{equation}
\mathcal{R}_{\mu\nu}\mathcal{R}^{\mu\nu} \to \tilde{\mathcal{R}}_{\mu\nu}\tilde{\mathcal{R}}^{\mu\nu}=\frac{\mathcal{R}_{\mu\nu}\mathcal{R}^{\mu\nu}}{\Omega_{k}^{4}(x)}=\frac{\mathcal{R}_{\mu\nu}\mathcal{R}^{\mu\nu}}{\mathcal{\mathcal{R}}^{2h(k)-2}h^{2}(k)}=
\begin{cases}
      \mathcal{R}_{\mu\nu}\mathcal{R}^{\mu\nu}, & \text{for}\ h(k) \rightarrow 1,\\
      \frac{1}{16}, & \text{for}\ h(k) \rightarrow 2.
\end{cases}
\end{equation}

\noindent Again, for any power $n$ of $\left(\mathcal{R}_{\mu\nu}\mathcal{R}^{\mu\nu}\right)$ the result will be a constant in the high-energy limit. For $h(k) \rightarrow 2$ it is straightforward to show that $\left(\mathcal{R}_{\mu\nu}\mathcal{R}^{\mu\nu}\right)^{n} \to 1/16^{n}$.

Perhaps the most widely used curvature invariant is the Kretschmann scalar $K$, where

\begin{equation}
K \vcentcolon =\mathcal{R}_{\mu\nu\rho\sigma}\mathcal{R}^{\mu\nu\rho\sigma}=\mathcal{C}_{\mu\nu\rho\sigma}\mathcal{C}^{\mu\nu\rho\sigma}+2\mathcal{R}_{\mu\nu}\mathcal{R}^{\mu\nu}-\frac{1}{3}\mathcal{R}^{2},
\end{equation}

\noindent and $C_{\mu\nu\rho\sigma}$ is the Weyl curvature tensor. The Weyl curvature tensor measures the initial tidal tistortion of a small sphere released from rest and allowed to free-fall under gravity~\cite{Penrose91}. Experiments, specifically measurements of the Cosmic microwave background (CMB) radiation and the spectral index, indicate that the early universe was very close to being homogenous and isotropic. As pointed out by Penrose~\cite{Penrose91}, had it been exactly homogenous and isotropic the Weyl curvature would have been exactly zero. In fact, observations are entirely consistent with the exact result $C_{abcd}=0$ at the big bang~\cite{Penrose91}. Assuming a vanishing Weyl tensor, the Kretschmann scalar in the very early universe simplifies to

\begin{equation}
K=2\mathcal{R}_{\mu\nu}\mathcal{R}^{\mu\nu}-\frac{1}{3}\mathcal{R}^{2}.
\end{equation}

\noindent It is now easy to show that $K$ transforms under~(\ref{weyl1}) as 








\begin{equation}
K \rightarrow \tilde{K}=\frac{K}{\Omega_{k}^{4}(x)}=\frac{2\mathcal{R}_{\mu\nu}\mathcal{R}^{\mu\nu}-\frac{1}{3}\mathcal{R}^{2}}{\mathcal{R}^{2h(k)-2}h(k)^{2}}=
\begin{cases}
      K, & \text{for}\ h(k) \rightarrow 1,\\
      \frac{1}{24}, & \text{for}\ h(k) \rightarrow 2.
\end{cases}
\end{equation}

\noindent It is important to stress that in the Palatini formalism it is possible that the Kretschman scalar may not in fact be a scalar, as indicated by the recent work of Ref.~\cite{Bejarano:2019zco}. Furthermore, we have assumed that the Weyl tensor is exactly zero at the putative big bang singularity~\cite{Penrose91}. 

Even disregarding the Kretschmann scalar, we have still shown that the high-energy limit of this model contains no curvature singularity in any of the curvature invariants $\mathcal{R}$, $\mathcal{R}^{2}$ or $\mathcal{R}_{\mu\nu}\mathcal{R}^{\mu\nu}$, to any power $n$. This suggests that in this scenario there may be no curvature singularity associated with the big bang. 

In the case of singularities in Schwarzschild spacetime, $R$ and $R_{\mu\nu}$ are identically zero and so cannot be used as curvature invariants. If, however, the Kretschman scalar is a valid curvature invariant in the Palatini formalism, contrary to the findings of Ref.~\cite{Bejarano:2019zco}, then since $K\propto r^{-6}$ as a function of the Schwarzschild coordinate $r$ and $K \to K \Omega_{r}^{-4}(x)$ then this implies that $\Omega_{r}(x)$ must behave like $\Omega_{r}(x) \sim r^{-\frac{3}{2}}$ in order to recover a constant in the high-energy limit. In close proximity to the putative singularity at $r=0$, classical proper distance $d_{0}$ scales non-trivially as $d_{0}\propto r^{\frac{3}{2}}$~\cite{Adeifeoba:2018ydh}. This suggests the Weyl factor should scale like $\Omega_{d_{0}}(x) \sim 1/d_{0}$, or conversely like $\Omega_{k}(x) \sim k$, at small proper distances or high energies~\cite{Adeifeoba:2018ydh,Coumbe:2018myj}. 

Note that t'hooft has already shown that conformal invariance formally removes the singularity inside black holes~\cite{tHooft:2016uxd}, which in combination with this work suggests a theory that is conformally symmetric at high energies is free from any type of singularity. 





\end{section}


\begin{section}{Discussion}

\begin{subsection}{Is the model viable?}

In this subsection we discuss whether or not AWIG is wrong in an obvious way, and hence whether it can even be considered as a possible alternative to general relativity. There are a number of criteria that must be satisfied for this model to be even considered viable. The proposed model must~\cite{Sotiriou:2008rp}:
\begin{enumerate}[label=(\roman*)., start=1]
\item Be consistent with experiment.
\end{enumerate}

Whether or not our model satisfies this most important of criteria depends on the exact functional form of $h(k)$ and the energy scale at which it exhibits a measurable deviation from unity. Given the limiting values of $h(k)$ assumed in our model (see Eq.~(\ref{a1})) we may equivalently write $h(k)=1+g(k)$, with $g(k)=0$ in the IR and $g(k)=1$ in the UV. Since $g(k)$ must be a dimensionless function of the scale $k$ we might guess a functional form such as

\begin{equation}\label{a3}
g(k)= \sum_{n=1}^{\infty}c_{n} \left(\frac{k}{k_{*}}\right)^{n},
\end{equation}

\noindent where $k_{*}$ is the energy scale at which deviations from general relativity become significant. Gravitational theories of the type $f(R)=R^{1+\delta}$, where $\delta \in \mathbb{R}$, have been experimentally constrained by cosmological and solar system tests, albeit within the metric formalism~\cite{Clifton:2005aj}. Firstly, a stable matter-dominated period of evolution requires that $\delta \geq 0$ or $\delta < -1/4$~\cite{Clifton:2005aj}. Secondly, such a modified action changes the expansion rate during primordial nucleosynthesis altering the predicted abundance of light elements, which can be used to place the constraint $-0.0064 <\delta <0.0012$ for a specific range of the baryon-to-photon ratio~\cite{Olive:1999ij,Clifton:2005aj}. The tightest bounds come from the perihelion precession of Mercury, which constrains $\delta$ at the level $\delta< 7.2\times 10^{-19}$~\cite{Clifton:2005aj}, meaning we must have $g(k_{sol}) <7.2\times 10^{-19}$ at the scale $k_{sol}$ associated with such solar system experiments. Since $k_{*}$ could in principle be arbitrarily large there can be no conflict with such experiments, that is until the function $g(k)$ is known.  

The 2018 CMB anisotropy measurements by the Planck Satelite provide important constraints on early universe cosmology and cosmic inflation~\cite{Akrami:2018odb}. In particular, the scalar spectral index was determined to be $n_{S}=0.9649\pm 0.0042$ with a $68\%$ CL, and in combination with results from BICEP2 and the Keck Array the tensor-to-scalar ratio has the bound $r<0.064$~\cite{Akrami:2018odb}. Applying these results to the analysis of Ref.~\cite{Rinaldi:2014gua} suggests the inflationary phase of the universe favours an action $f(R)=R^{2-\epsilon}$, where $\epsilon$ quantifies small ($|\epsilon|\ll 1$) deviations from exact scale invariance. Recent cosmological data also suggests an empirical relationship between the scalar spectral index $n_{S}$ and the number of e-foldings $N$ of the form $n_{S}-1\simeq 2/N$. In Ref.~\cite{Chiba:2018cmn} this relationship was used to reconstruct the functional form of $f(R)$ gravity, with the result that $f(R)\to R^{2}$ in the large $R$ limit. This result also supports an inflation-era action that is near scale-invariant. The analysis of Ref.~\cite{Rinaldi:2014gua} and~\cite{Chiba:2018cmn} was performed within the metric formalism of $f(R)$ gravity, not the Palatini formalism, and so it only strictly supports a globally scale invariant action at high energies. However, since global scale invariance is a special type of local scale invariance these findings are at least consistent with our model.

It has been argued that Palatini $f(R)$ gravity may be in conflict with the standard model of particle physics~\cite{Flanagan:2003rb}. This argument is based on the observation that $\mathcal{R}$ is an algebraic function of the traced energy tensor $T$ for a given $f(\mathcal{R})$, as is evident from the traced field equations~(\ref{traceFE})~\cite{Sotiriou:2008rp}. It is argued that since the energy tensor and its trace contain first order derivatives of the matter fields, terms such as $f'(\mathcal{R})$ appearing in the field equations will contain two or more derivatives of the matter fields. In general relativity, there is no such problem, since in this case $f'(R)$ is a constant and so its field equations are only first order in matter fields. We point out that in our model $F(\mathcal{R})$ is a purely algebraic function of $T$ only at a constant energy scale $k$. Moreover, as long as AWIG only deviates from general relativity in the vicinity of the ultraviolet limit, at the much lower energy scales relevant to the standard model of particle physics $F'(\mathcal{R})$ will be experimentally indistinguishable from a constant, thereby evading any such conflict.

\begin{enumerate}[label=(\roman*)., start=2]
\item Reproduce known cosmological phases.
\end{enumerate}

Observations support a standard model of cosmology with four phases: an early phase of accelerated expansion, followed by a radiation dominated phase, a matter dominated phase, and the present phase of accelerated expansion. The need to account for this phase structure has produced numerous competing explanations. A major motivation for considering Palatini gravity is that it appears particularly well-suited to this task~\cite{Fay:2007gg,Amendola:2006kh}. 



Palatini $f(R)$ gravity is equivalent to general relativity with a cosmological constant when the trace of the energy tensor vanishes, namely as $T \to 0$. Equation~(\ref{traceFE}) tells us that $T\to 0$ when $h(k)\to 2$. Since the energy scale $k$ of the very early universe was extremely high one expects $h(k)\to 2$ in this regime, and so our model becomes equivalent to general relativity with a cosmological constant in this case, presumably yielding an early period of rapid cosmic expansion~\cite{Edery:2019txq}. 



Following this early period of expansion, the energy scale $k$ will decrease sending $h(k)\to 1$, and hence reproducing the dynamics of standard general relativity without a cosmological constant. The exact dynamics between these initial and final states depends on the interplay between the matter density and the effective cosmological constant. If at earlier times the matter density dominates over the effective cosmological constant then we should expect to recover the usual radiation and matter dominated phases. If at later times the matter density goes to zero faster than the effective cosmological constant, then at some point the accelerating effect of a residual cosmological constant will begin to dominate over the decelerating effect of matter density, which is consistent with the present phase of accelerated expansion. The cosmological dynamics described by this scenario is essentially equivalent to the standard cosmological model of general relativity plus an assumed cosmological constant. The difference, however, is that Palatini $f(R)$ gravity does not assume a cosmological constant from the outset, it is naturally induced by the dynamics of the theory. 

Starting from an action of the type $f(\mathcal{R})=\mathcal{R}^{h(k)}$ it can be shown that~\cite{Allemandi:2004ca,Carroll:2003wy}

\begin{equation}
w_{eff}=\frac{1}{h(k)}\left(1+w\right)-1, \qquad h(k)\neq 2,
\end{equation}   

\noindent where $w$ is the equation of state of a perfect fluid $w\equiv p/ \rho$ and $w_{eff}$ is the equivalent modified quantity in our model. The only cosmological phase in which $h(k)$ will significantly deviate from unity is in the very early universe, thus $w_{eff}\simeq w$ at all but the very earliest moments. Therefore, cosmological dynamics should be very similar to standard general relativity including dark energy in the form of a cosmological constant, since $w_{eff}\simeq w$ in all phases. 

\begin{enumerate}[label=(\roman*)., start=3]
\item Have the correct weak field limit at low energies.
\end{enumerate}

The model we propose reduces to general relativity in the low-energy limit, a theory which is known to have the correct weak field limit. Mathematically, satisfying this criteria requires that $f''(R)\geq 0$, where $f''(R)$ denotes the second order partial derivative of $f(R)$ with respect to $R$~\cite{Sotiriou:2008rp}. Since AWIG has the specific form $F(\mathcal{R})=\mathcal{R}^{h(k)}$ then $F''(\mathcal{R})=\mathcal{R}^{h(k)-2}h(k)\left(h(k)-1\right)$ and so as $h(k)\to 1$ (low energy regime) we have $F''(\mathcal{R})=0$. Thus, AWIG exactly saturates the bound demarking the correct weak field, just as normal general relativity does. 

\begin{enumerate}[label=(\roman*)., start=4]
\item Be classically and semiclassically stable.
\end{enumerate}

$f(R)$ theories does not suffer from an Ostragadsky instability, since these theories uniquely violate Ostragadsky's assumption of non-degeneracy~\cite{Woodard:2006nt}. This establises $f(R)$ models as the only metric-based, local and potentially stable modifications of gravity~\cite{Woodard:2015zca,Woodard:2006nt}. Theories for which $f''(R)\geq 0$ avoid the Dolgov-Kawasaki instability~\cite{Dolgov:2003px}, which is the case in our model since $h(k)\geq 1$. In general, the Dolgov-Kawasaki instability only occurs in the metric formalism and so is not relevant to our model. In fact, it has been explicitly demonstrated that the Dolgov-Kawasaki instability cannot occur in any formulation of Palatini gravity, due to the fact that in these theories the scalar field is not dynamical~\cite{Sotiriou:2008rp,Sotiriou:2006sf}.

\begin{enumerate}[label=(\roman*)., start=5]
\item Contain no unphysical ghost modes.
\end{enumerate}

No type of $f(R)$ theory contains ghost modes~\cite{Sotiriou:2008rp}. Mathematically, satisfying this criteria requires $f'(R)> 0$, so that the effective gravitational coupling $G_{eff}=G/f'(R)$ cannot be negative and the conformal transformation of the metric $g_{\mu\nu} \to g_{\mu\nu} f'(R)$ is properly defined. In our model $F'(\mathcal{R}) > 0$ as long as $h(k) > 0$, which it clearly is since we assume from the outset that $1 \leq h(k) \leq 2$. 

\begin{enumerate}[label=(\roman*)., start=6]
\item Have a well-posed Cauchy problem.
\end{enumerate}

In the high-energy limit ($h(k)\to 2$), our model certainly has a well-posed Cauchy problem since in this case it is equivalent to general relativity with a cosmological constant~\cite{Olmo:2008nf,Edery:2019txq}. In the low-energy limit ($h(k)\to 1$) our model recovers general relativity with a vanishing cosmological constant, and so the Cauchy problem is once again well-posed in this regime, even when including low-energy matter content. However, in between these two extremes ($1< h(k)< 2$) it is less clear whether our model satisfies this criteria. In the range $1< h(k)< 2$, our model contains no vacuum solution ($T\neq 0$) and in this case a well-posed Cauchy problem seems to require rather specific conditions that may be hard to satisfy with generic matter fields~\cite{WaldBook}, although this point remains an open issue and requires a careful examination using specific matter fields~\cite{Sotiriou:2008rp}.\\ 

To summarise this section, it appears that the model proposed is not obviously wrong. It seems to satisfy many of the standard viability criteria, although it must be significantly developed in order to provide a more complete assesment of its viability. A detailed analysis, especially one involving the inclusion of specific matter fields, would help decide this issue, particularly with regard to the Cauchy problem.

\end{subsection}


\begin{subsection}{Quantum gravity}

Here we briefly discuss how this proposal relates to certain aspects of quantum gravity. 

\begin{subsubsection}{Conformal anomaly}

The idea that the ultimate laws of nature contain no fixed scale is an aesthetically appealing one. Yet it is of little practical use without a symmetry-breaking mechanism that allows scales to emerge at lower energies.

The local scale symmetry described here is a symmetry of classical spacetime. Quantization typically breaks any exact conformal symmetry present in a classical action, a phenomenon known as the conformal anomaly~\cite{Mukhanov:2007zz}. A defining characteristic of a conformal field theory is that its traced energy tensor vanishes, namely $T=0$. For example, in our model, the symmetry of local scale invariance is only exact in the high energy limit $h(k)\to 2$, for which $T=0$ identically. However, in a quantum field theory on a curved background $T$ does not generally vanish, although in flat space it still does~\cite{Mukhanov:2007zz}. This is a serious challenge to any conformally invariant classical theory. This problem is substantial and beyond the scope of the present work, but nevertheless, it is important to point out. 

\end{subsubsection}

\begin{subsubsection}{Dimensional reduction}
  A wide variety of approaches to quantum gravity have reported a dynamical reduction in the number of spacetime dimensions in the high energy limit (see Refs.~\cite{Carlip:2017eud,Carlip:2019onx} for a review). Remarkably, these independent approaches almost universally find the same ultraviolet dimensionality of two. Since in $d$-dimensions the gravitational coupling $\kappa$ of standard general relativity has a canonical mass dimension of $[\kappa]=2-d$, an ultraviolet dimensionality of two will result in a dimensionless gravitational coupling. Thus, standard general relativity would become renormalizable in the ultraviolet, at least at the level of power-counting. Moreover, in $2$-dimensional spacetime gravity becomes locally scale-invariant, behaving in the same way as a conformal field theory~\cite{Banks:2010tj,Shomer:2007vq,Falls:2012nd}. Unfortunately, dimensional reduction has unphysical implications including an energy-dependent speed of light and deformations or violations of Lorentz invariance~\cite{Sotiriou:2011mu,Amelino-Camelia:2013tla}. Such predictions have become heavily constrained by several recent experiments~\cite{Coumbe:2015aev}. For example, the \emph{Fermi} space telescope has definitively shown the speed of light to be independent of energy up to $7.62 E_{P}$ (with a $95 \%CL$) and $\simeq 4.8 E_{P}$ (with a $99 \%CL$)~\cite{Vasileiou:2013vra} for linear dispersion relations, where $E_{P}$ is the Planck energy.

Similarly, AWIG also has a dimensionless gravitational coupling in the high-energy limit (see section~\ref{action}), is renormalizable~\cite{Stelle:1977nt}, and by construction behaves in the same way as a conformal field theory at high energies. However, AWIG is diffeomorphism invariant, appears to satisfy the Einstein equivalence principle, and its matter action should trivially reduce to that of special relativity locally~\cite{Sotiriou:2008rp}. Therefore, the scale-dependent $F(\mathcal{R})$ action described by AWIG seems to have the same positive physical features as dimensional reduction, such as renormalizability and high-energy conformal scaling, without the negative aspects such as an energy-dependent speed of light and deformations or violations of Lorentz invariance~\cite{Coumbe:2015zqa,Coumbe:2015bka,Coumbe:2015aev,Coumbe:2018myj}.  

\end{subsubsection}
\end{subsection}
\end{section}


\begin{section}{Conclusions}

  Demanding a classical 4-dimensional diffeomorphism invariant theory of gravity that: (i) is renormalizable, (ii) evades Ostragadsky's no-go theorem, and (iii) is locally scale-invariant, uniquely leads to pure $f(R)=R^{2}$ gravity in the Palatini formalism. We also demand that this theory reduce to general relativity in the low-energy limit, for which $f(R)=R$. In order to satisfy these demands, we propose a novel theory of Palatini gravity in which $f(R)=R^{h(k)}$, where $h(k)$ is a dimensionless function of the energy scale $k$ that smoothly interpolates between $h(k)=2$ in the UV and $h(k)=1$ in the IR. To the best of my knowledge, this is the first time such a model has been proposed.

The model we propose is applied to the well-known problem of curvature singularites in classical general relativity. We have demonstrated that the proposed model contains no cosmological curvature singularities in $\mathcal{R}$ or $\mathcal{R}_{\mu\nu}\mathcal{R}^{\mu\nu}$, at any order $n$. We have also shown that the Kretschmann scalar $K=\mathcal{R}_{\mu\nu\rho\sigma}\mathcal{R}^{\mu\nu\rho\sigma}$ contains no cosmological curvature singularity, assuming $K$ is a valid curvature invariant in Palatini gravity~\cite{Bejarano:2019zco}.

Our model appears to pass a number of key viability criteria, at least qualitatively. However, a more complete quantitative assesment is required in order to properly test the models viability, a study that will particularly benefit from the inclusion of specific matter fields.

\end{section}

\vspace{0.5cm}
\noindent \emph{Acknowledgments}: I wish to acknowledge support from the Danish Research Council grant \emph{Quantum Geometry}.


\bibliographystyle{unsrt}
\bibliography{Master}

\end{document}